\documentclass[aps,pra,twocolumn,superscriptaddress]{revtex4-1}
\usepackage{graphicx}
\usepackage{amsmath,amssymb,amsfonts}
\usepackage{comment}
\usepackage{float}
\usepackage{placeins}

\usepackage[colorlinks,
	linkcolor=blue,
	citecolor=blue,
	urlcolor=blue]{hyperref}

\usepackage{cleveref}

\newcommand{\bom}[1]{{\mbox{\boldmath $#1$}}}
\def\chim{\chi_{\mathrm{m}}}
\def\kappain{\kappa_{\mathrm{in}}}
\def\kappaout{\kappa_{\mathrm{out}}}
\def\tin{t_{\mathrm{in}}}
\def\tout{t_{\mathrm{out}}}
\def\rin{r_{\mathrm{in}}}
\def\rout{r_{\mathrm{out}}}

\def\omegaopt{\omega_{\mathrm{opt}}}
\def\omegadip{\omega_{\mathrm{dip}}}
\renewcommand{\t}[1]{\mathrm{#1}}

\begin{document}

\title{Improving force sensitivity by amplitude measurement of light reflected from a detuned optomechanical cavity}

\author{Kentaro Komori}
\email[]{komori.kentaro@jaxa.jp}
\affiliation{Department of Physics, University of Tokyo, Bunkyo, Tokyo 113-0033, Japan}
\affiliation{Institute of Space and Astronautical Science, Japan Aerospace Exploration Agency, Sagamihara, Kanagawa 252-5210, Japan}
\author{Takuya Kawasaki}
\affiliation{Department of Physics, University of Tokyo, Bunkyo, Tokyo 113-0033, Japan}
\author{Sotatsu Otabe}
\affiliation{Department of Physics, Tokyo Institute of Technology, Meguro, Tokyo 152-8550, Japan}
\author{Yutaro Enomoto}
\affiliation{Department of Applied Physics, School of Engineering, The University of Tokyo, 7-3-1 Hongo, Bunkyo-ku, Tokyo 113-8656, Japan}
\author{Yuta Michimura}
\affiliation{Department of Physics, University of Tokyo, Bunkyo, Tokyo 113-0033, Japan}
\author{Masaki Ando}
\affiliation{Department of Physics, University of Tokyo, Bunkyo, Tokyo 113-0033, Japan}

\date{\today}

\begin{abstract}
The measurement of weak continuous forces exerted on a mechanical oscillator
is a fundamental problem in various physical experiments. It is fundamentally
impeded by quantum back-action from the meter used to sense the displacement
of the oscillator.
In the context of interferometric displacement measurements, we here propose and
demonstrate the working principle of a scheme for coherent back-action 
cancellation. By measuring the amplitude quadrature of the light
reflected from a detuned optomechanical cavity inside which a stiff optical spring is generated, back-action can be cancelled in a narrow band of frequencies.
This method provides a simple way to improve the sensitivity in experiments limited by quantum back-action without injection of squeezed light or stable homodyne readout.
\end{abstract}

\maketitle

\section{Introduction}

Precise mechanical sensing of forces has a long history, and a rich future. 
In the past, that pursuit has been exemplified by tests of Newtonian gravity \cite{Cavendish1798, Kapner2007, Schlamminger2008, Matsumoto2019, Westphal2021}, nanomechanics-based 
force microscopy \cite{Rug92,Rug04}, and gravitational-wave detection using kilogram-scale
test masses \cite{Abbott2016, Abbott2017}. 
The integration of nanoscale mechanical oscillators with optical cavities ---
within the field of cavity optomechanics \cite{Chen2013,Aspelmeyer2014} 
--- has opened the possibility of addressing a new set of questions through precision mechanical sensing. 
Examples include tests of gravitational effects in quantum mechanics \cite{Diosi1984, Penrose1996, Bassi2017, Marshall2003, Balushi2018}, 
tests of fundamental decoherence phenomena \cite{Bassi2013, Diosi2015, Nimmrichter2014,Carlesso2016,Helou2017,Komori2020}, and dark matter detection \cite{Carney2021,MonMoore20,Michimura2020,Morisaki2021,ManWil21}. The common denominator in all these quests is the broadband and precise measurement of forces acting on
mechanical oscillators.

The estimation of weak continuous forces is fundamentally limited by quantum noise. When optical fields are used
to measure the displacement of a mechanical force transducer, vacuum fluctuations in the former produce a
fluctuating back-action force that can obscure an external force \cite{Caves1981}. Such quantum back-action can be reduced by injection of light whose quadratures are squeezed in a frequency-dependent
fashion \cite{Kimble2001,Buonanno2001,Buonanno2002,Yap2020,Tse2019,Acernese2019}, 
or by employing ponderomotively generated squeezed light \cite{Sud17,Kamp17,Mason19,Yu2020}. 
In the context of force detection, 
the standard quantum limit (SQL) in the free-mass regime \cite{Khalili2012} $S_F^\t{SQL}(\omega) = 2\hbar m \omega^2$ 
decreases with decreasing frequency. Yet the opportunity for high precision broadband force sensing is thwarted by
the inability to generate squeezed light at low frequencies and/or phase noise in homodyne detection (as 
required for schemes that rely on ponderomotive squeezed light).

In this paper, we theoretically describe and experimentally demonstrate the principle of a technique that can circumvent
both these technical problems. In particular, we show that direct amplitude detection of the light
reflected from a detuned optomechanical system can realize quantum noise cancellation around the optical
spring frequency. 
Working in reflection has the fundamental advantage of the better sensitivity with an over-coupled cavity and beating the SQL, in contrast to similar schemes that operate in transmission~\cite{Cripe2020}.
Our scheme also does not require squeezed light or phase-stable homodyne detection to produce
force sensitivities beyond the SQL.
We also demonstrate the principle behind this scheme by showing that classical intensity noise in the laser
used to probe the optomechanical system is suppressed in a manner consistent with theory.

\section{Principle of quantum noise cancellation in amplitude readout}
\label{sec:method}

\begin{figure}
\centering
\includegraphics[width=\hsize]{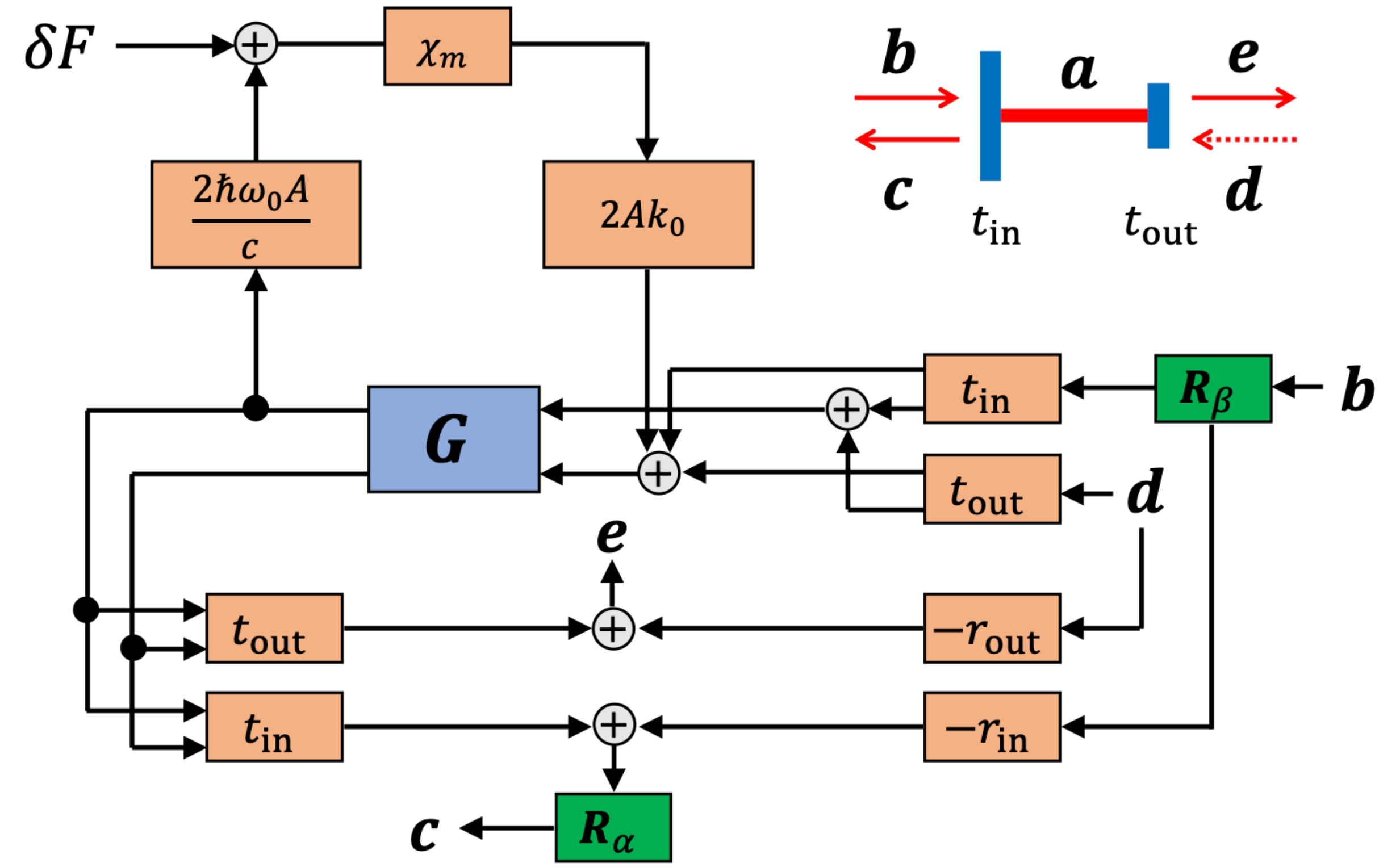}
\caption{
Block diagram for the input output relations. The brown blocks mean the scalar transfer functions such as the mechanical susceptibility and the transmission of the mirrors. The blue block shown as $G$ is the cavity matrix, and the green blocks are the rotation matrix to convert the input and reflection phase to the intra-cavity phase. At the right top area, we define the quadratures of the light in this paper.
}
\label{figure1}
\end{figure}

We consider a Fabry-Perot cavity with a mechanically compliant end-mirror of transmissivity $\tout$ and reflectivity $\rout = \sqrt{1-\tout^2}$, and a fixed input mirror of transmissivity $\tin$ and reflectivity $\rin = \sqrt{1-\tin^2}$. All optical loss in the system is modeled as being due to the non-zero transmissivity of the end-mirror. The linear dynamics of the system, driven by radiation pressure forces in the cavity, can be understood using a simple feedback picture as shown in Fig.~\ref{figure1}.
Adopting the two-photon formalism \cite{CavSch85,Corbitt2005},
we consider each optical field in the inset as being composed of a pair of quadratures; 
e.g. $\bom{a} = (a_1 \; a_2)^t$, where $a_1$ ($a_2$) is an amplitude (phase) quadrature of the intra-cavity light. Other fields are defined as shown in the inset: $\bom{b}$ ($\bom{c}$) is going into (out from) the input mirror, and $\bom{d}$ ($\bom{e}$) is going into (out from) the end mirror. The displacement caused by the external force $\delta F$ can be read out by $\bom{c}$ or $\bom{e}$ with the carrier. The Cavity amplitude is represented as $A=\sqrt{2P/(\hbar \omega_0)}$, where $P$ is the intra-cavity power, $\hbar$ and $\omega_0$ are Dirac constant and the angular resonant frequency of laser light, respectively. The Speed and wave number of light are defined as $c$ and $k_0=\omega_0 /c$. A matrix of the cavity amplification $\bom{G}$ is given by
\begin{equation}
     \bom{G} = \frac{c}{2L} \frac{1}{(\kappa - i \omega)^2 + \Delta^2}
\begin{pmatrix}
     \kappa - i\omega & -\Delta \\
     \Delta & \kappa - i\omega
\end{pmatrix}
,
\end{equation}
where $L$ is the cavity length, $\kappa = (t_{\mathrm{in}}^2+t_{\mathrm{out}}^2)c/(4L)$ is the total cavity decay rate, and $\Delta$ is the cavity detuning. In the detuned cavity, the carrier phase of the input, intra-cavity (or transmitted), and reflected light are different so that phase rotation should be taken into account. The phase rotation matrix is shown as
\begin{equation}
     \bom{R}_{\theta} = 
\begin{pmatrix}
     \cos \theta & - \sin \theta \\
     \sin \theta & \cos \theta
\end{pmatrix}
,
\end{equation} 
and $\alpha$ ($\beta$) is the phase difference between intra-cavity and reflected (input and intra-cavity) light.

Unlike the usual treatment in nano-optomechanics \cite{Bott12}, we specialize to the case of macroscopic
optomechanical systems where the mirror's motional frequency are so low that its utility as a broadband force
transducer necessitates measurement frequencies above its resonance. In this case the mirror motion is
in the free-mass regime, i.e. its displacement response to a force is, $\chim \simeq -1/(m\omega^2)$, where $m$ is the 
effective mass of the mirror.

\begin{figure}
\centering
\includegraphics[width=\hsize]{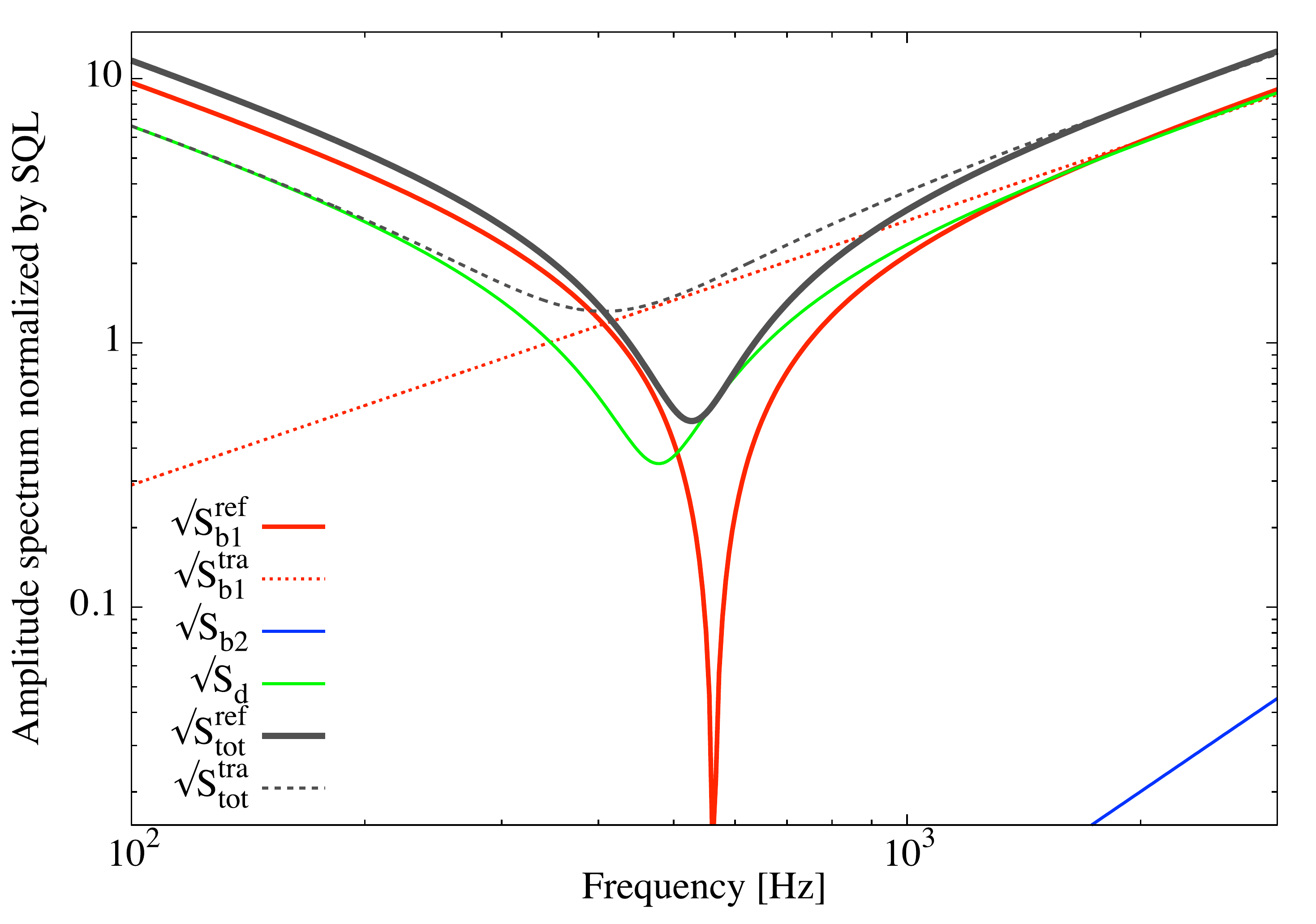}
\caption{
Amplitude spectra of amplitude and phase fluctuations in reflection and transmission measurement. Contributions of the input amplitude fluctuation at the reflection ($S_{b_1}^{\mathrm{ref}}$), the input phase fluctuation ($S_{b_2}$), and the vacuum from the output ($S_{d}$) are shown as red, blue, and green lines, respectively. Total sensitivities at the reflection ($S^{\mathrm{ref}}_\mathrm{tot}$) is represented as a black solid line. The input amplitude ($S_{b_1}^{\mathrm{ref}}$) and the total ($S^{\mathrm{tra}}_\mathrm{tot}$) at the transmission are plotted by red and black dotted lines.
}
\label{figure2}
\end{figure}

Here we focus on measurement of the amplitude quadratures at reflection and transmission, in other words $c_1$ and $e_1$. The sensitivity on these amplitude measurements of the reflection and transmission normalized by the SQL can be separated by the contributions of $\bom{b}$ and $\bom{d}$ as
\begin{gather}
    S^{\mathrm{ref}}_\mathrm{tot} = \varepsilon_{1} S^{\mathrm{ref}}_{b_1} + \varepsilon_{2} S_{b_2} + S_{d}, \\
    S^{\mathrm{tra}}_\mathrm{tot} = \varepsilon_{1} S^{\mathrm{tra}}_{b_1} + \varepsilon_{2} S_{b_2} + S_{d},
\end{gather}
where $\varepsilon_{1}$ and $\varepsilon_{2}$ are relative shot noise levels of the amplitude and phase of the input light.
These spectra are given by (see details in Appendix.~\ref{sec:detailcal})
\begin{gather}
\label{eq:specrefb1}
    S^{\mathrm{ref}}_{b_1} = \frac{ \left( \kappa^2 + \Delta^2 \right) \left\{ \Delta \iota - \left[ \left( \kappa - 2\kappain \right)^2 + \Delta^2 \right] \omega^2 \right\}^2 }{16 \iota \kappain \left(\kappa - \kappain \right)^2 \Delta^2 \omega^2}, \\
\label{eq:spectrab1}
    S^{\mathrm{tra}}_{b_1} = \frac{\kappain \left( \kappa^2 + \Delta^2 \right) \omega^2 }{\iota \Delta^2}, \\
\label{eq:specb2}
    S_{b_2} = \frac{\kappain \omega^4}{\iota (\kappa^2 + \Delta^2)}, \\
    S_{d} = \frac{ \left[ \Delta \iota - \left( \kappa^2 + \Delta^2 - 2\kappa \kappaout \right) \omega^2 \right]^2 }{4 \iota \kappaout \Delta^2 \omega^2} + \frac{\kappaout \omega^2 }{\iota},
\end{gather}
where
\begin{equation}
    \iota = \frac{4Pk_0}{mL}.
\end{equation}
The input and output coupler are given by $\kappain = \tin^2 \kappa /(\tin^2+\tout^2)$ and $\kappaout = \kappa - \kappain$ respectively. The amplitude spectra, normalized by the SQL, from the input fluctuation and the vacuum from the output port are plotted in Fig.~\ref{figure2} with the total ones. Here we assume that the input fluctuation is at the vacuum level, $\varepsilon_{1} = \varepsilon_{2} = 1$. Parameters are as follows: the laser wave length of $\lambda=1064$\,nm, $L=10$\,cm, $m=10$\,mg, $\kappa/(2\pi)=0.25$\,MHz, $\kappain / \kappa=0.8$, $\Delta = \kappa / \sqrt{3}$, and $P=1$\,W.

The sensitivity at the reflection reaches below unity, which means beating the SQL, due to the dip in $S^{\mathrm{ref}}_{b_1}$. This noise reduction occurs at the frequency where the amplitude fluctuation of the direct reflection and that of the cavity leakage are cancelled each other. The latter is dominant in the over-coupled cavity at high frequencies, while the former is larger at low frequencies because of the optical spring. The dip frequency is given by
\begin{equation}
    \omegadip = \sqrt{ \frac{\Delta \iota}{ \left( \kappa - 2\kappa_{\mathrm{in}} \right)^2 + \Delta^2 } },
\end{equation}
and it is always larger than the resonant frequency of the optical spring (Eq.~\ref{eq:optspring}). As for the input phase fluctuation shown by the blue line in Fig.~\ref{figure2}, the contribution to the total noise is much smaller than the others so that it is negligible. In this paper, we demonstrate the dip-shaped spectrum of the amplitude fluctuation, which is the most critical to the better force sensitivity at the quantum level.

We discuss difference between the reflection and transmission measurement. As shown by the dotted lines in Fig.~\ref{figure2}, the noise from the input amplitude fluctuation at the transmission is smaller that at the reflection ($S^{\mathrm{tra}}_{b_1} < S^{\mathrm{ref}}_{b_1}$), resulting in slightly better total sensitivity at low frequencies. This type of back action evasion was experimentally demonstrated by the previous work~\cite{Cripe2020}. Comparing to the transmission, the reflection measurement has an advantage of achieving better force sensitivity ultimately. The sensitivity of the typical over-coupled cavity is better at the dip frequency in the reflection measurement than at low frequencies in the transmission measurement. In addition, beating the SQL is an unique benefit of the reflection measurement.

\section{Experimental proof-of-principle}

\begin{figure}
\centering
\includegraphics[width=\hsize]{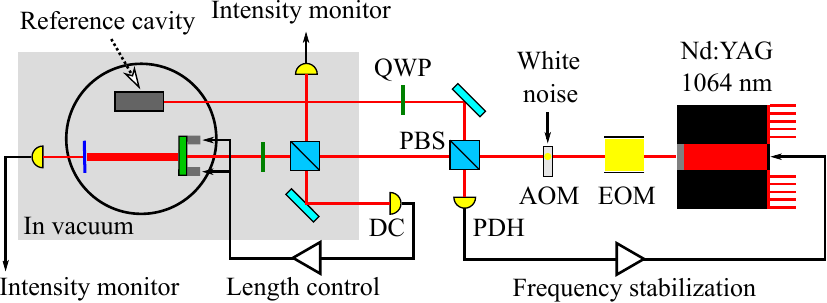}
\caption{
Schematic overview of the experimental setup. We use the output of a 1064\,nm laser. Reflected light from the main cavity is split and detected by a polarizing beam splitter (PBS), a quarter wave plate (QWP), and a DC photodiode. The error signal controls the cavity length with the feedback to the input mirror with the coil-magnet actuator. The classical radiation pressure and intensity noises are excited by the injection of the white noise to an acousto-optic modulator (AOM). The tranmission of the cavity is monitored to estimate the cavity detuning during the measurement. An electro-optic modulator (EOM) generates the phase modulation for the frequency stabilization with a reference cavity and a Pound-Drever-Hall (PDH) error signal.
}
\label{figure3}
\end{figure}

We demonstrate the proof-of-principle of the technique mentioned above in the experiment depicted in Fig.~\ref{figure3}.
A 11-cm linear cavity consists of a 8 mg end-mirror (0.5\,mm thick with a diameter 3\,mm) suspended by a single carbon fiber (6\,$\mu$m thick and 2\,cm long), and a much heavier (60 g) input mirror. 
The radii of curvature of the mirrors are 10\,cm, shorter than the cavity length of 11\,cm; this autonomously stabilizes the cavity
against radiation-pressure torque instabilities~\cite{Kawasaki2021}.
In order to realize the optical spring, the laser is blue-detuned from cavity resonance. The required error
signal to stabilize the detuning is derived from the power reflected from the cavity away from resonance. The
error signal is compared against a DC reference, which is then fed back to actuate on coil-magnet actuators on the input mirror that controls the cavity length.
By changing the DC reference, we measure several sensitivities to the external force acting on the test mass with different detuning. The transmission of the cavity is monitored to estimate the detuning during the measurement.
The cavity is driven by the input power $P_{\mathrm{in}} \simeq 4.7$\,mW of 1064-nm light from a Nd:YAG laser derived at a beam splitter.
The finesse is measured to be $\mathcal{F}=(3.0\pm 0.3)\times 10^3$, resulting in the intra-cavity power $P \sim 5$\,W.

The system is not in a regime where quantum radiation pressure fluctuations dominate the motion of the end mirror.
Nevertheless, the principle underlying coherent radiation pressure noise cancellation can
be demonstrated on calibrated classical radiation pressure noise impressed on the input light. 
The injection is performed by adding the white noise to an acousto-optic modulator (AOM) the input light passes through, so that the mg-scale test mass is driven by the classical radiation pressure noise. It is confirmed by coherence between the error signal and the intensity noise, which is taken by the photo detector for the light picked off just before the cavity. The injected noise is so large ($\sqrt{\varepsilon_{1}} \sim \mathcal{O}(10^3)$) that the coherence is measured to be almost unity at all frequencies.

In order to observe the classical radiation pressure fluctuation with better signal-to-noise ratio, the laser frequency is stabilized by a reference cavity (4.4\,cm long,
with a finesse $6.4\times 10^4$). The reflected light from the reference cavity, whose phase is modulated by an electro-optic modulator (EOM), is used as the Pound-Drever-Hall (PDH) error signal. Produced feedback signal actuates a laser PZT and stabilizes the frequency noise. The reference cavity is co-located with the experimental cavity, on a vacuum vibration isolation platform. The pressure is kept around 100\,Pa to avoid the coupling of acoustic noise and simultaneously make the cavity locked more easily due to the residual gas damping.

\begin{figure}
\centering
\includegraphics[width=\hsize]{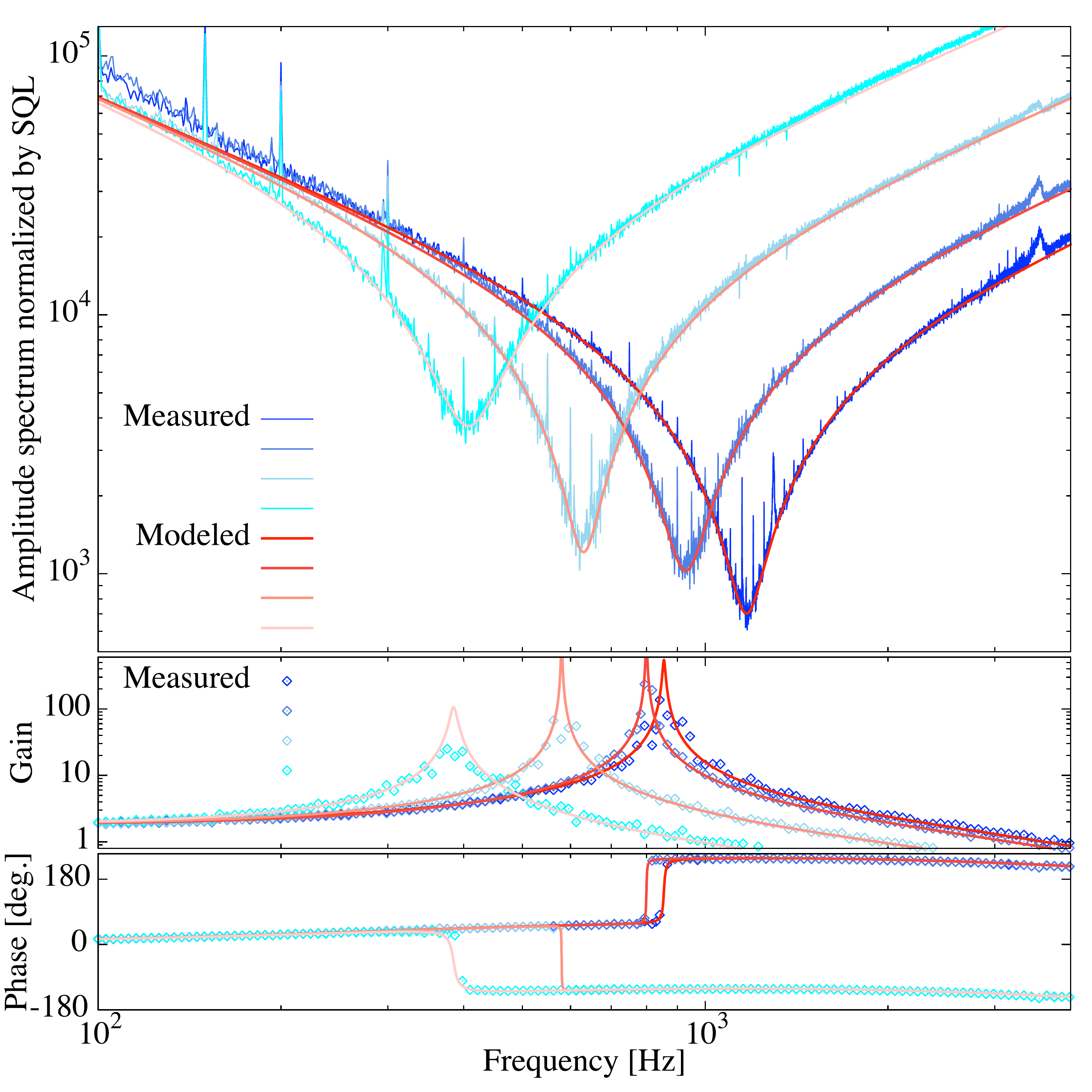}
\caption{
Demonstration of the dip-shaped spectrum from the amplitude fluctuation of the input light. The amplitude spectrum normalized by the SQL is shown at the upper panel, and the openloop transfer function is shown at the middle and bottom panels. The experiment is performed in four different detuning, and the results (blue curves and points) are fitted by the modeled curves (red lines).}
\label{figure4}
\end{figure}

\section{Result and Discussion}

Our experiments are performed with four different detuning keeping the constant input power. The result of the four measurements is shown in Fig.~\ref{figure4}. At the upper panel, the spectra normalized by the SQL are plotted with the modeled curves. Those are calibrated from the error signal to the force sensitivity by the transfer function of the openloop and the filter for the length control. The gains and phases of the openloop transfer function are plotted at the middle and bottom panels with the modeled curves. At the resonant frequency, the phase is advanced (delayed) in the two measurements with higher (lower) intra-cavity power, since the negative damping of the optical spring does (not) overwhelm the residual gas damping. The negative damping is compensated by the electrical feedback loop.

In the case of the conventional phase measurement in this setup, we should see the flat force sensitivity without the normalization because of the classical radiation pressure noise. In our experiment, the dip-shaped reduction of the noise is clearly observed by the amplitude measurement of the reflection. Without the detuning fluctuation during the measurement, the sensitivity with injection of the white intensity noise has the spectrum of
\begin{equation}
    \sqrt{S^{\mathrm{ref}}_{b_1} (\omega_{\mathrm{dip,m}})} \propto \frac{\left| \omega_{\mathrm{dip,m}}^2 - \omega^2 \right|}{\omega_{\mathrm{dip,m}}^2},
\end{equation}
where $\omega_{\mathrm{dip,m}}$ is the measured dip frequency. In practice, the detuning is changing so that the dip gets thicker. In order to estimate the dip frequency with the error, we assume that the dip frequency distributes as Gaussian where the central frequency is $\omega_{\mathrm{dip,m}}$ with the standard deviation of $\delta \omega$. The modeled curve is generated by averaging the two distributed spectra because of the Gaussianity, $\sqrt{(S^{\mathrm{ref}}_{b_1} (\omega_{\mathrm{dip,m}} + \delta \omega) + S^{\mathrm{ref}}_{b_1} (\omega_{\mathrm{dip,m}} - \delta \omega))/2}$, and the fitting is performed by the two parameters and the overall factor. In this way, for instance, we estimate the dip frequency in the measurement with the highest power as $\omega_{\mathrm{dip}}/(2 \pi) = 1180 \pm 70$\,Hz.

\begin{figure}
\centering
\includegraphics[width=\hsize]{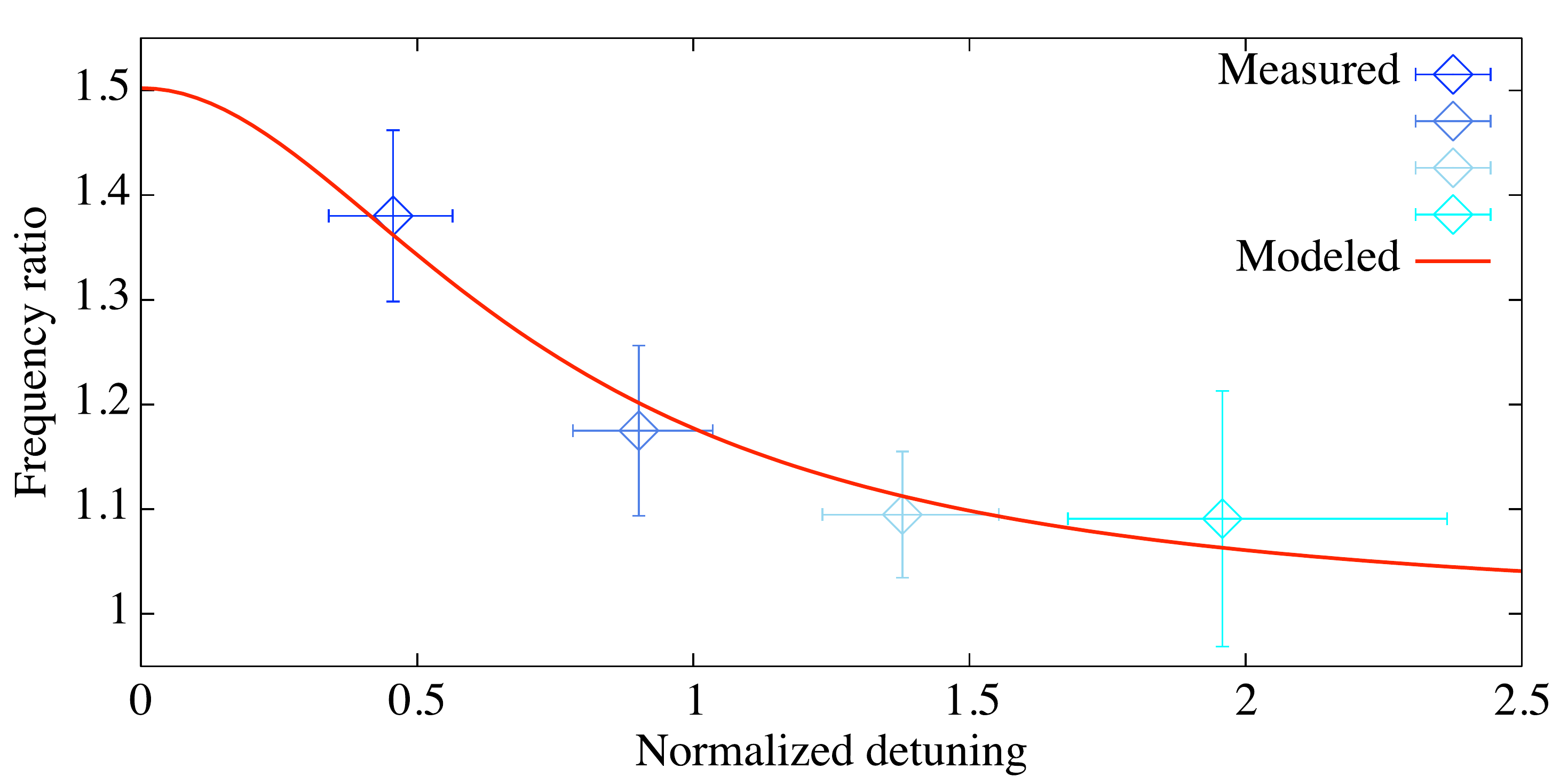}
\caption{
Ratios between optical spring frequencies and dip frequencies. The four blue dots with errors represent the results with the different detuning. Each colored dot corresponds to the estimation from the same colored spectrum in Fig.~\ref{figure4}. The red line shows the modeled curve fitted to the measured data.}
\label{figure5}
\end{figure}

In Fig.~\ref{figure5}, we show ratios between optical spring frequencies and dip frequencies at the four different detunings. The ratio is a solid indicator to evaluate the system since it is not affected by uncertainty of the intra-cavity power. The mean values and errors of the ratio are calculated from the modeling of the measured spectra in the previous paragraph. The detunings are estimated by comparing the transmission of the cavity during the measurement with the maximum output in the cavity scan. The errors of the detunings come from the residual fluctuations of the transmission power. When we make the modeled curve of the ratio, the effect of the mode mismatch between the cavity TEM00 mode and the input beam must be taken into account. The mismatched light is directly reflected from the cavity, and contributes as the sensing noise which has a dip at the optical spring frequency (see details in Appendix~\ref{sec:modemis}). Due to this effect, the measured dip frequency is smaller than that in the perfect mode matching. The mode matching ratio is measured to be $\eta = 92\%$. The transmission of the input mirror is estimated to be $\kappain / \kappa = 0.81$ by the fitting.

\section{Conclusion}
The optomechanical cavity can be used for the broadband and precise measurement of forces acting on mechanical oscillators, such as the quantum decoherence phenomena and the dark matter interaction, but the sensitivity is fundamentally limited by the quantum noise. We theoretically show that the force sensitivity of the test mass trapped by the optical spring can be improved as the dip by measuring the amplitude of the light reflected from the detuned cavity. We experimentally demonstrate the dip-shaped improvement with the mg-scale suspended mirror by adding the intensity modulation to the light. This method does not require additional setup for the squeezed light or homodyne measurement. We conclude that the amplitude measurement of the reflection gives a simple way to improve the sensitivity even beyond the quantum limit.

\section{Acknowledgement}
We thank V. Sudhir for help with the manuscript and fruitful discussions. We would like to acknowledge R. Yamada, B. Lane, N. Mavalvala, and K. Somiya for discussions. We also thank OptoSigma company for producing the mg-scale curved mirror. This work was supported by JST CREST Grant No. JPMJCR1873.

\appendix
\section{Detailed calculations}
\label{sec:detailcal}

In this section, we describe the detailed calculation to derive the force sensitivities shown in Sec.~\ref{sec:method}. The intra-cavity field is enhanced by the cavity amplification synchronizing with the cavity displacement. The loop gain of of the detuned cavity is given by products of all blocks in the optomechanical loop as
\begin{equation}
    \bom{\mathcal{A}} = 
    \begin{pmatrix}
    0 & 0 \\
    - \kappa_0 & 0
    \end{pmatrix}
    \bom{G},
\end{equation} 
where $\kappa_0 = -8 \chim P k_0 / c$. The cavity response to the input fields can be written as
\begin{align}
\bom{H} &= \bom{G} \left( \bom{I}-\bom{\mathcal{A}} \right)^{-1} \nonumber \\
&= -\frac{c \omega^2}{2L\mathcal{M}}
\begin{pmatrix}
\kappa - i\omega & -\Delta \\
\Delta - \iota / \omega^2 & \kappa - i\omega
\end{pmatrix}
,
\end{align}
where $\bom{I}$ is an unit matrix, and
\begin{equation}
    \mathcal{M} = \omega^2 \left[ ( \omega + i \kappa )^2 - \Delta^2 \right] + \Delta \iota.
\end{equation}
The resonant frequency of the optical spring can be calculated by $\mathrm{Re} (\mathcal{M}) = 0$, that is,
\begin{equation}
\label{eq:optspring}
    \omegaopt = \sqrt{\frac{\Delta \iota}{\kappa^2 + \Delta^2}}.
\end{equation}
The input-output relations for quadratures of the reflection and the transmission are expressed as
\begin{align}
\begin{pmatrix}
c_1 \\
c_2
\end{pmatrix}
&= \bom{R}_{\alpha} \left\{ \left[ t_{\mathrm{in}}^2 \bom{H} - r_{\mathrm{in}} \bom{I} \right] \bom{R}_{\beta}
\begin{pmatrix}
b_1 \\
b_2
\end{pmatrix}
\right. \nonumber \\ &\left. + t_{\mathrm{in}} t_{\mathrm{out}} \bom{H}
\begin{pmatrix}
d_1 \\
d_2
\end{pmatrix}
+ 2\chim Ak_0 t_{\mathrm{in}} \bom{H}
\begin{pmatrix}
0 \\
\delta F
\end{pmatrix}
\right\} ,
\end{align}
and
\begin{align}
\begin{pmatrix}
e_1 \\
e_2
\end{pmatrix}
&= \left[ t_{\mathrm{out}}^2 \bom{H} - r_{\mathrm{out}} \bom{I} \right]
\begin{pmatrix}
d_1 \\
d_2
\end{pmatrix}
\nonumber \\ &+ t_{\mathrm{in}} t_{\mathrm{out}} \bom{H} \bom{R}_{\beta}
\begin{pmatrix}
b_1 \\
b_2
\end{pmatrix}
+ 2\chim Ak_0 t_{\mathrm{out}} \bom{H}
\begin{pmatrix}
0 \\
\delta F
\end{pmatrix}
.
\end{align}
The phases between different carriers are given by
\begin{gather}
    \alpha = \arctan \left(- \frac{\Delta}{2\kappain - \kappa} \right), \\
    \beta = \arctan \left(- \frac{\Delta}{\kappa} \right), \\
    \gamma = \arctan \left(- \frac{2\kappain \Delta}{2\kappa \kappain - \kappa^2 - \Delta^2} \right),
\end{gather}
where $\gamma = \alpha + \beta$ is the phase difference between the input and the reflected light. In the limit where the transmissivity of the mirrors are much smaller than 1, we can safely approximate $\tin^2 \simeq 4L \kappain /c $, $\tout^2 \simeq 4L \kappaout /c $, and $\rin \simeq \rout \simeq 1$.

We should note the phase rotation of the input and reflection against the intra-cavity light. Light with career has different phases at three points, input, intra-cavity, and reflection so that $\bom{R}_{\alpha}$ and $\bom{R}_{\beta}$ are considered. Unlike the career light, however, the input from the end mirror is a vacuum field, so the quadrature rotation of $\bom{d}$ does not have to be considered in calculating total quantum noise caused by that vacuum. Also, it should be noted that there is no quadrature rotation in transmission measurement because the intra-cavity and transmission carriers have the same phase.

We focus on the amplitude fluctuation of the reflection and transmission. The amplitude of the reflection can be calculated as
\begin{equation}
    c_1 = \chi_{\mathrm{ref}} \left( \delta F + \xi_{b_1} b_1 + \xi_{b_2} b_2 + \xi_{d_1} d_1 + \xi_{d_2} d_2 \right),
\end{equation}
where
\begin{equation}
    \chi_{\mathrm{ref}} = 2 \chim A k_0 \tin \left( H_{12} \cos \alpha - H_{11} \sin \alpha \right),
\end{equation}
\begin{multline}
    \xi_{b_1} = \Bigl[ \tin^2 \bigl( H_{11} \cos \gamma + H_{12} \cos \alpha \sin \beta \\
    - H_{21} \sin \alpha \cos \beta \bigr) -\rin \cos \gamma \Bigr] / \chi_{\mathrm{ref}},
\end{multline}
\begin{multline}
    \xi_{b_2} = \Bigl[ \tin^2 \bigl( - H_{11} \sin \gamma + H_{12} \cos \alpha \cos \beta \\
    + H_{21} \sin \alpha \sin \beta \bigr) + \rin \sin \gamma \Bigr] / \chi_{\mathrm{ref}},
\end{multline}
\begin{gather}
    \xi_{d_1} = \frac{\tout}{2 \chim A k_0} \frac{H_{11} \cos \alpha - H_{21} \sin \alpha }{H_{12} \cos \alpha - H_{11} \sin \alpha}, \\
    \xi_{d_2} = \frac{\tout}{2 \chim A k_0}.
\end{gather}
The components of matrix $\bom{H}$ is expressed as $H_{ij}$ ($H_{22} = H_{11}$). The amplitude of the transmission is given by
\begin{equation}
    e_1 = \chi_{\mathrm{tra}} \left( \delta F + \eta_{b_1} b_1 + \eta_{b_2} b_2 + \eta_{d_1} d_1 + \eta_{d_2} d_2 \right),
\end{equation}
where
\begin{gather}
    \chi_{\mathrm{tra}} = 2 \chim A k_0 \tout H_{12}, \\
    \eta_{b_1} = \frac{\tin}{2 \chim A k_0} \frac{H_{11} \cos \beta + H_{12} \sin \beta}{H_{12}}, \\
    \eta_{b_2} = \frac{\tin}{2 \chim A k_0} \frac{ -H_{11} \sin \beta + H_{12} \cos \beta}{H_{12}}, \\
    \eta_{d_1} = \frac{\tout^2 H_{11} - \rout}{\chi_{\mathrm{tra}}}, \\
    \eta_{d_2} = \frac{\tout}{2 \chim A k_0}.
\end{gather}

In our calculation, frequencies we are interested in are much smaller than the cavity line width ($\kappa \gg \omega$) so that the second order term of $(\omega / \kappa)^2$ can be ignored. Also, the ;aser power is assumed to be not so large that the optical spring is much smaller than the cavity pole, $\omegaopt \ll \kappa$. Since the power and cross spectrum of the vacuum field is given by Kronecker delta ($S_{a_i a_j} = \delta_ {ij}$), the power spectra of the force noise corresponding to each input field at the amplitude measurement of the reflection are given by
\begin{align}
    S^{\mathrm{ref}}_{F,b_1} &= \left| \xi_{b_1} \right|^2 \nonumber \\
    &= \frac{ \left( \kappa^2 + \Delta^2 \right) \left\{ \Delta \iota - \left[ \left( \kappa - 2\kappain \right)^2 + \Delta^2 \right] \omega^2 \right\}^2 }{16 \iota \kappain \left(\kappa - \kappain \right)^2 \Delta^2 \omega^2} S_F^\t{SQL},
\end{align}
\begin{gather}
    S^{\mathrm{ref}}_{F,b_2} = \left| \xi_{b_2} \right|^2 = \frac{\kappain \omega^4}{\iota (\kappa^2 + \Delta^2)} S_F^\t{SQL}, \\
    S^{\mathrm{ref}}_{F,d_1} = \left| \xi_{d_1} \right|^2 = \frac{\left[ \Delta \iota - \left( \kappa^2 + \Delta^2 - 2\kappa \kappaout \right) \omega^2 \right]^2 }{4 \iota \kappaout \Delta^2 \omega^2} S_F^\t{SQL}, \\
    S^{\mathrm{ref}}_{F,d_2} = \left| \xi_{d_2} \right|^2 = \frac{\kappaout \omega^2 }{\iota} S_F^\t{SQL}.
\end{gather}
Those at the transmission measurement are
\begin{gather}
    S^{\mathrm{tra}}_{F,b_1} = \left| \eta_{b_1} \right|^2 = \frac{\kappain \left( \kappa^2 + \Delta^2 \right) \omega^2 }{\iota \Delta^2} S_F^\t{SQL}, \\
    S^{\mathrm{tra}}_{F,b_2} = \left| \eta_{b_2} \right|^2 = \frac{\kappain \omega^4}{\iota (\kappa^2 + \Delta^2)} S_F^\t{SQL}, \\
    S^{\mathrm{tra}}_{F,d_1} = \left| \eta_{d_1} \right|^2 = \frac{\left[ \Delta \iota - \left( \kappa^2 + \Delta^2 - 2\kappa \kappaout \right) \omega^2 \right]^2 }{4 \iota \kappaout \Delta^2 \omega^2} S_F^\t{SQL}, \\
    S^{\mathrm{tra}}_{F,d_2} = \left| \eta_{d_2} \right|^2 = \frac{\kappaout \omega^2 }{\iota} S_F^\t{SQL}.
\end{gather}
We note that the only difference between the reflection and transmission measurement comes from the noise of the input amplitude fluctuation, and the others are the same. By normalizing these force spectra by the SQL and incorporating the two spectra of the vacuum from the output ($S_d = S_{d_1} + S_{d_2}$), we get the expressions in the main text.

The input phase fluctuation $b_2$ includes a frequency noise of the laser light. Defining the power spectrum of the angular frequency noise as $S_{\mathrm{freq}}$, the relative shot noise level is calculated by dividing the equivalent phase noise of $S_{\mathrm{freq}}/\omega^2$ by a normalization factor $\hbar \omega_0 /(2P_{\mathrm{in}})$, where $P_{\mathrm{in}}$ is the input power. The relative shot noise level of the frequency noise can be written as
\begin{equation}
\varepsilon_{2} = \frac{2P_{\mathrm{in}}}{\hbar \omega_0 \omega^2} S_{\mathrm{freq}},
\end{equation}
and the intra-cavity power is
\begin{equation}
    P = \frac{c \kappain}{L (\kappa^2 + \Delta^2)} P_{\mathrm{in}}.
\end{equation}
Thus, the displacement spectrum from the carrier phase fluctuation is given by
\begin{equation}
\chim^2 \varepsilon_{2} S_{b_2}  = \frac{L^2}{\omega_0^2} S_{\mathrm{freq}},
\end{equation}
which is a typical expression of the frequency noise.

\section{Effect of mode mismatch}
\label{sec:modemis}

In this section, we discuss the effect of mode mismatch on the dip frequency measured in the spectrum. We write the power of the TEM00 mode of the cavity and that of other modes as $P_{00} = \eta P_{\mathrm{in}}$ and $P_{\mathrm{mm}} = (1- \eta) P_{\mathrm{in}}$, respectively. The amplitude fluctuation of the reflection for the TEM00 mode is given by
\begin{multline}
    \chi_{\mathrm{ref}} \xi_{b_1} b'_1 \\
    = \sqrt{\frac{\kappa^2 + \Delta^2}{(\kappa - 2\kappain)^2 + \Delta^2}} \frac{\Delta \iota - \left[ (\kappa - 2\kappain)^2 + \Delta^2 \right] \omega^2}{\mathcal{M}} b'_1,
\end{multline}
where $b'_1$ is the classical amplitude noise, and the absolute value is proportional to $\sqrt{P_{00}}$. Noting that the reflected field of the TEM00 mode can be represented with the input field $|E_{00}| (\propto \sqrt{P_{00}})$ as
\begin{equation}
   |E_{\mathrm{ref},00}| = \sqrt{\frac{(\kappa - 2\kappain)^2 + \Delta^2}{\kappa^2 + \Delta^2}} |E_{00}|,
\end{equation}
The power fluctuation of the reflected TEM00 is approximately expressed as
\begin{align}
    \delta P_{\mathrm{ref},00} &\propto |E_{\mathrm{ref},00}| |\chi_{\mathrm{ref}} \xi_{b_1}| b'_1 \nonumber \\
    &\propto \frac{\Delta \iota - \left[ (\kappa - 2\kappain)^2 + \Delta^2 \right] \omega^2}{\Delta \iota - (\kappa^2 + \Delta^2) \omega^2} \eta P_\mathrm{in}.
\end{align}
On the other hand, the power of the mismatched light directly reflects from the cavity, and couples to the power fluctuation as the intensity noise,
\begin{equation}
    \delta P_{\mathrm{mm}} \propto (1 - \eta) P_{\mathrm{in}}.
\end{equation}
Therefore, the total fluctuation of the reflection power is given by $\delta P_{\mathrm{ref}} = \delta P_{\mathrm{ref},00} + \delta P_{\mathrm{mm}}$, and the measured dip frequency in the spectrum of $\delta P_{\mathrm{ref}}$ is
\begin{equation}
    \omega_{\mathrm{dip,m}} = \sqrt{\frac{\Delta \iota}{\kappa^2 + \Delta^2 - 4\kappain (\kappa - \kappain) \eta}}.
\end{equation}

\bibliography{paper}
\bibliographystyle{apsrev.bst}
\end{document}